\documentclass[aps,twocolumn,showpacs,floatfix]{revtex4}

  \usepackage{graphics,graphicx,epsfig}

\begin{document}

\title{Demonstration of the spatial separation of the entangled quantum
side--bands of an optical field}
\author{E.H.Huntington, G. N. Milford, C.
Robilliard}
\affiliation{Centre for Quantum Computer Technology, School of
Information Technology and Electrical Engineering, University
College, The University of New South Wales, Canberra, ACT, 2600}
\author{T. C. Ralph}
\affiliation{Centre for Quantum Computer Technology,  Department of
Physics, The University of Queensland,  St Lucia QLD 4072 Australia}
\author{O. Gl\"{o}ckl, U. L.
Andersen, S. Lorenz, G. Leuchs}
\affiliation{Max--Planck Forschungsgruppe, Institut f\"ur Optik, Information und
Photonik, \\Universit\"at Erlangen--N\"urnberg, G\"unther--Scharowsky--Str. 1 /
Bau 24, 91058 Erlangen, Germany }

\begin{abstract}
     Quantum optics experiments on ``bright" beams typically probe
     correlations between side-band modes.  However the extra degree of
     freedom represented by this dual mode picture is generally
     ignored.  We demonstrate the experimental operation of a device
which can be
     used to separate the quantum side-bands of an optical field. We
     use this device to explicitly demonstrate the quantum entanglement between
     the side-bands of a squeezed beam.

\end{abstract}

\maketitle

\vspace{10 mm}

Early on in the discussion of squeezed light it was realized that a single
mode description is often not adequate. The spectral structure of 
squeezed light
with its frequency side-bands is essential when analyzing specific Fourier
components of the field fluctuations \cite{Caves82,Unruh83,Caves85,Yurke85}.
 From a theoretical point of view it became
clear that the two side-bands of a squeezed beam carry quantum correlated noise
\cite{Gea87,Caves82,Caves85}. Later it was realized that squeezing
across two distinct spatial modes led to entanglement and
tests of the Einstein Podolsky Rosen (EPR) Gedankenexperiment using
such two mode squeezed light were proposed \cite{Graham84,Reid89} and
demonstrated
\cite{Kimble92,zhang00EPR,Silberhorn01}. Since then EPR entanglement has been
recognized as a basic resource of continuous variable quantum
information protocols \cite{cvbook}. Two mode squeezing involves
pair-wise correlations between the four frequency side-band modes of the two
beams.  Recently it has been suggested that the spectral side-band
correlations of single mode squeezed light can be transferred to
entanglement between two spatial modes \cite{hunt02,zhang03}.
The entanglement thus produced would be of a quite different character
to that produced by two mode squeezing as it would involve a pair wise
correlation between only a single side-band on each of the beams.

In this letter we demonstrate experimentally for the
first time the production of this new type of entangled light by
separating the quantum side-bands of a single spatial mode into
two separate spatial beams. We first demonstrate the basic effect
using continuous wave coherent states (experiment type A) and then demonstrate
entanglement production from pulsed
squeezed light (experiment type B). As well as the fundamental interest of
demonstrating this paradigm of quantum optics explicitly,
the techniques described here
represent a new tool in the analysis and manipulation of quantum
optical fields \cite{hunt04,merolla} and produce entanglement useful for
quantum information applications.
Note that the ability to detect a single side-band
may also open new possibilities for astronomy \cite{Townes}.

  \begin{figure}[htbp]
\begin{center}
  \includegraphics[width=8cm]{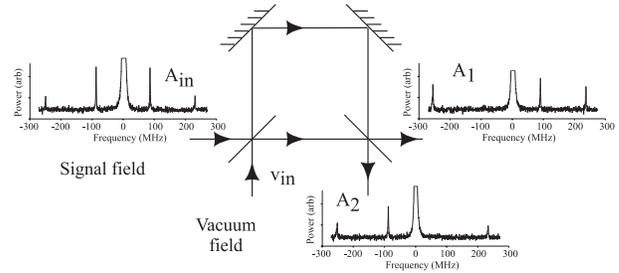}
   \caption{ \label{schemFBS}
A schematic diagram of the system along with spectra illustrating
successful operation at the classical level.  All the spectral measurements were
made with a scanning confocal Fabry-Perot cavity \cite{kogelnik} with
a Free-Spectral Range (FSR) of $500$MHz and a linewidth of
approximately $2$MHz.  The truncated carrier is shown at zero
frequency, the phase modulation side-bands are at $\pm 90.5$MHz and
residual mode mismatch peaks (with less than 1\% power) are at $\pm 250$MHz.  }
\end{center}
   \end{figure}

Fig.  \ref{schemFBS} illustrates our experimental set-up.  This figure
also shows sketches of input and output frequency spectra indicating
the desired operation of the system.  The system is essentially a
Mach-Zehnder interferometer whereby the path length for one of the
interferometer arms is much greater than for the other, thus
introducing a time delay between the two arms of $\tau$.  Previously
such unbalanced Mach-Zehnder interferometers (UMZI) have been used in
quantum optics experiments to filter spurious longitudinal lasing
modes \cite{yam?} and measure the phase quadrature of bright pulsed
beams \cite{glockl}.  Here we pick the path length difference such
that the quantum side-bands at a particular radio frequency are
decomposed into separate spatial beams.

All of the beamsplitters in the UMZI are assumed $50\%$ transmitting. 
In the Heisenberg picture, time varying fields are written in the
rotating frame as $\hat{{A}}(t)= \bar{{A}} + \delta \hat{{A}}(t)$ and
$\hat{{A}}(t)^{\dagger}= \bar{{A}}^{\star} + \delta
\hat{{A}}(t)^{\dagger}$ where the steady-state coherent amplitude of
the field is given by $\bar{{A}}$ and the time-varying component of
the field is given by the operator $\delta \hat{{A}}(t)$.  In Fourier
transform space, we may define operators for the quadrature amplitude
and phase fluctuations of a field as $\delta X^{+}(\omega) = \delta
A(\omega) + \delta A(-\omega)^{\dagger}$ and $\delta X^{-}(\omega) =
\imath(\delta A(\omega) - \delta A(-\omega)^{\dagger})$ respectively. 
The absence of hats indicates the Fourier transform.  Additionally we
have made use of the relation that $\delta \hat{{A}}(t) \rightarrow
\delta A(\omega) \Rightarrow \delta \hat{{A}}(t)^{\dagger} \rightarrow
\delta A(-\omega)^{\dagger}$ to find the relevant creation operators
in the frequency domain \cite{glauber}.  The spectral variances
$V^{\pm}$ of all fields are found from $\langle |\delta X^{\pm}
(\omega)|^{2} \rangle $.

The annihilation operators for the UMZI inputs (outputs) at a Fourier
frequency $\omega$ relative to the carrier are denoted $\delta
A_{in}(\omega)$ and $\delta v_{in}(\omega)$ ($\delta A_{1}(\omega)$
and $\delta A_{2}(\omega)$) - see Fig.  \ref{schemFBS}.  The outputs
of the UMZI are given in terms of the inputs as
\begin{eqnarray}
        \label{FBS1} \delta A_{1,\phi}(\omega)&=&\frac{1}{2} \left[ \delta
A_{in}(\omega)
        \left(1- e^{\imath \phi}e^{\imath \omega \tau} \right) \right.
\nonumber
        \\
        && \left. + \imath
        \delta v_{in}(\omega) \left( e^{\imath \phi} e^{\imath \omega \tau}+1
        \right) \right] \\
        \label{FBS2} \delta A_{2,\phi}(\omega)&=&\frac{1}{2} \left[ \imath
\delta A_{in}(\omega) \left( e^{\imath
        \phi}e^{\imath \omega \tau} + 1 \right) \right. \nonumber
        \\
        && \left.+ \delta v_{in}(\omega) \left(
        e^{\imath \phi} e^{\imath \omega \tau}-1 \right) \right]
        \end{eqnarray}
\noindent where $\phi=2m \pi+\omega_{0} \tau =\omega_{0} \tau $ is
the phase shift acquired by a field at the carrier frequency $\omega_{0}$. Let
us focus on the specific frequency $\Omega$ such that $\Omega \tau=\pi/2$.
Choosing $\phi=+\pi/2$, the outputs of the UMZI are
  \begin{eqnarray}
        \delta A_{1,+\pi/2}(\Omega)&=&\delta A_{in}(\Omega), \space
        \delta A_{1,+\pi/2}(-\Omega)= \imath \delta
v_{in}(-\Omega)
        \nonumber \\
        \delta A_{2,+\pi/2}(\Omega)&=&-\delta v_{in}(\Omega), \space
        \delta A_{2,+\pi/2}(-\Omega)= \imath \delta
A_{in}(-\Omega)
        \nonumber
        \end{eqnarray}
For $\phi=-\pi/2$ the $\pm\Omega$ terms are interchanged.

Hence we find that the positive and negative frequency components of
the input field can be separated into spatially distinct beams.  If
the UMZI is locked to $-\pi/2$ the upper and lower side-bands will
exit from the opposite ports.  This behaviour has been confirmed
experimentally and is indicated in Fig \ref{schemFBS}.  The spectra
illustrating successful operation at the classical level are the
results of measurements of a continuous wave Nd:YAG laser (Innolight
Mephisto 500) with $90.5$MHz phase modulation side-bands sent as an
input to the UMZI (experiment type A).  The UMZI had a path length
difference of $0.83$m and fringe visibility of $98\%$.  The DC power
at output $A_{1}$ was used to lock the interferometer.  Spectral
measurements of the field $A_{2}$ were made for $\phi= + \pi/2$ and
for $\phi= - \pi/2$.  The second of these is shown on the $A_{1}$
output to illustrate schematically the desired operation of the
device.  The spatial separation of the upper and lower side-bands is
clearly visible.

\begin{figure}[htbp]
\begin{center}
  \includegraphics[width=7.8cm]{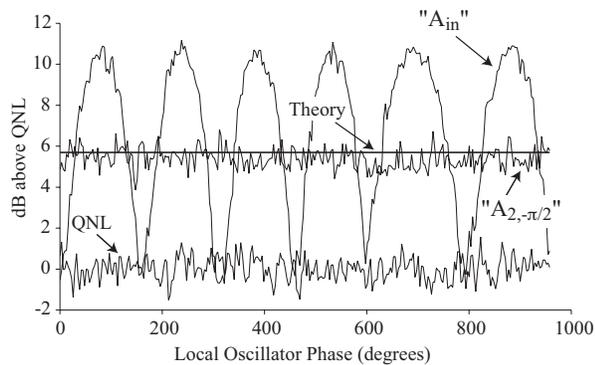}
\caption{ \label{homod}
Homodyne measurements of: the Quantum Noise Limit for the local
oscillator power (labelled QNL); the input to the UMZI (``${\rm
A_{in}}$''), one output of the UMZI when locked to $\phi=-\pi/2$
(``${\rm A_{2,-\pi/2}}$''); and the predicted output of the UMZI
based on the measured input amplitude and quadrature phase variances,
as well as the measured homodyne detection fringe visibilites
($0.89\%$
and $0.92\%$ for input and output respectively  \cite{bachor}) and UMZI fringe
visibility (``Theory'').}
\end{center}
\end{figure}

We now turn to non-classical effects.  The power spectrum of the input
fluctuations is given by
\begin{eqnarray}\label{vin}
V_{in}^{\pm}&=&\langle \delta A_{in}(\Omega)^{\dagger} \delta
A_{in}(\Omega) +
\delta A_{in}(-\Omega)^{\dagger} \delta A_{in}(-\Omega) \nonumber \\
&& \pm
\delta A_{in}(-\Omega) \delta A_{in}(\Omega) \pm
\delta A_{in}(-\Omega)^{\dagger} \delta
A_{in}(\Omega)^{\dagger}\rangle \nonumber \\ && +1
\end{eqnarray}
The final unit component in the sum is the vacuum noise.
Using the transfer relations for the UMZI ($\phi = +\pi/2$)
we find for the output fluctuations of the first beam
\begin{equation}\label{vout1}
      V_{out1}^{\pm}=\langle \delta A_{in}(\Omega)^{\dagger}
      \delta A_{in}(\Omega) \rangle +1
\end{equation}
where expectation values such as $\langle \delta v_{in}^{\dagger}
\delta v_{in} \rangle$, $\langle \delta A_{in}^{\dagger} \delta
v_{in}^{\dagger} \rangle$ and $\langle \delta A_{in} \delta v_{in}
\rangle$ are all zero.  $V_{out2}^{\pm}$ is similar to Eq.\ref{vout1}
with $\Omega$ replaced by $-\Omega$.  Thus we expect only the power of
the upper side-band to appear at the first output and only that of the
lower side-band to appear at the second.  Also we expect the spectra
to be independent of the local oscillator phase when probing the
side--bands with a homodyne detector.  If the input state is symmetric
(same power in the upper and lower side--bands) then the variances of
the amplitude and phase quadratures of both outputs of the UMZI when
locked to either $\pm \pi/2$ would be (normalized to quantum noise
limit, QNL, of one output)
\begin{equation}\label{vout}
      V_{out}= (V_{in}^{+}+V_{in}^{-}+2)/4
\end{equation}
Fig.~\ref{homod} shows the results of homodyne measurements of the
input and output of the UMZI for experiment type A at $90.5$MHz. 
Ref.~\cite{hunt04} outlines the approach taken to model mode-mismatch. 
To take an example, the intensity noise of either of the output fields
is $V_{out,mm}^{+} = (V_{in}^{+} + \eta_{mm} V_{in}^{-} + 3 -
\eta_{mm})/4$ where $\eta_{mm}$ is the mode-matching efficiency of the
UMZI given by the square of the fringe visibility.  There is good
agreement between the measured and predicted behaviour of the UMZI.

Of more interest is what happens if the input beam is squeezed.  From
Eq.~\ref{vout} we see that the noise of the output beam must be greater
than the QNL for all quadratures (as $V + 1/V$ is always greater than
2 for $V \ne 1$).  Indeed, from Eq.~\ref{vout1}, this shows that
squeezed beams have side-band photons.  So the output beams show excess
noise.  Now consider the correlations between the output beams.  If
amplitude quadrature measurements are simultaneously made on both
beams then the sum and difference photocurrent variances (normalised
to the QNL of both output beams) are given by
\begin{equation}\label{vadd+}
      V_{add}^{+}=(V_{in}^{+}+1)/2 \quad V_{sub}^{+}=(V_{in}^{- }+1)/2
\end{equation}
\noindent If the input beam is amplitude squeezed then Eq.~\ref{vadd+}
shows that amplitude measurements of the two beams are anti--correlated
to below the QNL. On the other hand if phase quadrature measurements
are simultaneously made on both beams then the normalised sum and
difference photocurrent variances are given by
\begin{equation}\label{vsub-}
      V_{add}^{-}=(V_{in}^{-}+1)/2 \quad V_{sub}^{- }=(V_{in}^{+}+1)/2
  \end{equation}
\noindent Thus phase quadrature measurements are correlated to below
the QNL. Sub--QNL correlations on both quadratures are the signature of
entanglement \cite{duan}, showing that the side--band entanglement has
been transferred to entanglement between spatially separated beams
\cite{hunt02}.

To demonstrate these quantum effect, we use an UMZI to separate the side--bands
of an amplitude squeezed input field. In this experiment (type B), we use a
commercially available pulsed Optical Parametric Oscillator pumped by
a mode locked Ti:Sapphire laser (both Spectra Physics). The Optical
Parametric Oscillator produces pulses of 130fs at a center wavelength of
1530nm and a repetition rate of 82MHz. The nonlinear Kerr effect
experienced by intense pulses in optical fibers (see e.~g. ref.~\cite{SIZ99}) is
used to generate non--classical states of light. During propagation of such
pulses through a fibre, a high degree of excess phase noise is introduced
mainly due to some classical noise effects. Employing an asymmetric fibre Sagnac
interferometer \cite{SCHM98}, amplitude squeezing can be produced, the amplitude
not being affected by excess noise.

For a pulsed laser beam the separation of side--bands can be performed only at
certain frequencies, as two conditions have to be fulfilled simultaneously. (1)
Two pulses have to overlap temporally giving a boundary condition for the path
length difference $\Delta L=cnT_{\rm rep}$ ($n$ is an integer number and $T_{\rm
rep}$ is the time between two pulses) and (2) A $\pi/2$--phase shift at the
measurement frequency $f_\mathrm{m}$ must be introduced, so that $\Delta L=c/(4
f_\mathrm{m})$. Possible measurement frequencies are therefore $f_{\rm m}=
1/(4nT_{\rm rep})$. At our repetition rate the arm length difference must be a
multiple of 3.66m, corresponding to the distance between two successive pulses.
The arm length difference is set to be 7.32m for measurements at $10.25$MHz.

\begin{figure}[htbp]
\begin{center}
  \includegraphics[width=7.8cm]{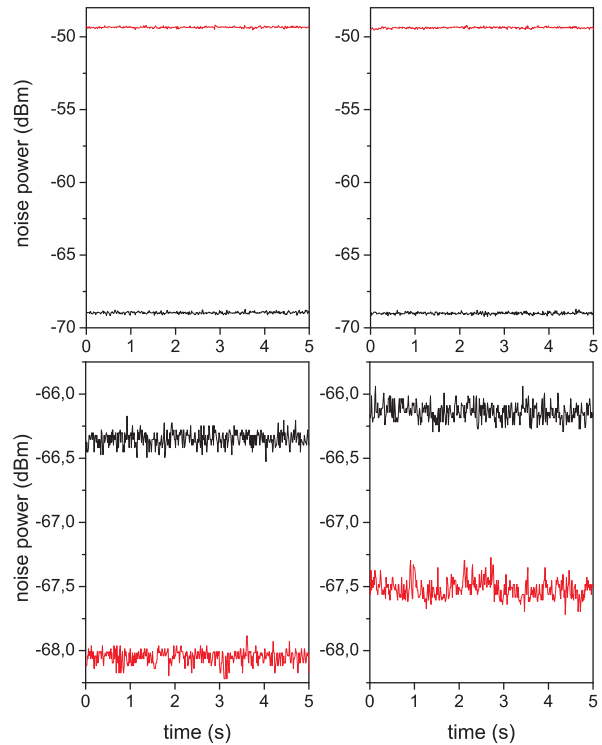}
   \caption{
\label{resultate} Characterization of the output beams of the UMZI when using
squeezed light. The upper two graphs show the amplitude noise level of the
individual beams (gray lines) compared to the respective quantum noise limit
(black lines). The lower left graph shows the correlations in the amplitude
quadrature (gray line) compared to the quantum noise limit (black line). The
lower right graph shows the correlations in the phase quadrature (gray line)
compared to the quantum noise limit (black line). All noise levels have been
detected using a pair of spectrum analyzers (HP 8590) at a measurement frequency
of $10.25$MHz, a resolution bandwidth of 300kHz and a video bandwidth of 30Hz.
All traces have been corrected for the electronic noise level, that was at about
$-77.9$dBm. }
   \end{center}
\end{figure}

We launched about $4$dB of amplitude squeezed light into the UMZI. A
visibility of 95\% was observed at the output to generate entanglement. On each
of the two output beams we performed an amplitude noise measurement (see the
upper two graphs in Fig.~\ref{resultate}). The amplitude noise of the output
beams (gray traces) is almost 20dB above the quantum noise limit (black traces).
This high noise level is an indication that strong correlations between pairs of
side--bands might be present. Note also that the squeezing is lost as a result
of the separation of side--bands. The high noise level is not only due to
antisqueezing, but also due to the high classical thermal phase noise of our
squeezed states.

To demonstrate that this
beam pair is indeed entangled, correlations below the quantum noise limit
between the two beams in the amplitude as well as in the phase quadrature must
be observed. While amplitude correlations are verified rather easily in
direct detection of the two output modes and subsequently correlating the
photocurrents, phase measurements are more involved. Having intense pulsed light
trains, neither homodyne detection (that might saturate our detectors due to the
high intensities involved) nor phase shifting cavities (with high requirements
on resonance conditions) could be easily employed. We therefore used an
interferometric scheme where two entangled beams interfere at yet another 50/50
beam splitter. Their relative phase is such that the output beams
(denoted a and b) of this interference have equal intensity \cite{Silberhorn01}.
Both output beams are then detected directly. The spectral component at
frequency $\omega$ of the sum and the difference of the photocurrents from
the two detectors yields signals that are proportional to the sum of the
amplitude and the difference of the phase quadratures of the two entangled beams
respectively
\begin{eqnarray} V(\delta n_\mathrm{a}(\omega)+\delta
n_\mathrm{b}(\omega))&=&V(\delta X_1^+(\omega)+\delta
X_2^+(\omega))=V_\mathrm{add}^+\nonumber \\
V(\delta n_\mathrm{a}(\omega)-\delta
n_\mathrm{b}(\omega))&=&V(\delta X_1^-(\omega)-\delta
X_2^-(\omega))=V_\mathrm{sub}^-.\nonumber
\end{eqnarray}
The measurement scheme is equivalent to a Bell state measurement on intense
light beams \cite{LEU99ZHA00}. It turned out that this type of measurement can
also be applied to the entangled beam pair generated by separating the
side--bands to check for the correlations in the amplitude and the phase
quadrature. In the verification experiment, we achieved a visibility of more
than 90\%. We observed about 1.6dB of correlations
below the quantum noise limit in the amplitude quadrature (lower left traces in
Fig.~\ref{resultate}) and about 1.4dB of correlations in the phase quadrature
(lower right traces in Fig.~\ref{resultate}).  This experiment shows that the quantum correlations between the
side--bands of the squeezed input beam have been successfully transferred to
quantum correlations between two spatially distinct outputs. The result clearly
shows the entanglement between the outputs, as both, non--classical correlations
in the amplitude as well as in the phase quadrature of the beam pair were
detected. Note that the degree of observed correlations of $1.6$dB for the
amplitude quadrature ($1.4$dB for the phase quadrature) agrees with the
prediction from theory. About $1.6$dB of correlations are expected when $4$dB of
input squeezing are used, as the correlations are degraded by the contribution
of the uncorrelated vacuum side--bands (see Eqs. \ref{vadd+} and 
\ref{vsub-}). To extract the full correlations of
$4$dB a pair of frequency shifted local oscillators would have to be applied as
proposed by Zhang \cite{zhang03}.

In summary, we have demonstrated the production of a new type of
entangled light via a device which can be used to separate the quantum
side-bands of an optical field.  This device is in essence an
unbalanced Mach-Zehnder interferometer (UMZI) where the path length
difference is determined by the frequency at which side-band
separation is required.  We have shown that the UMZI may be used to
spatially separate the positive and negative side-bands of a phase
modulated optical field.  Applying this to single mode squeezed light
produces spatially separated entangled beams.  Many landmark
experiments have investigated quantum properties with homodyne
detection, for example \cite{breit97,furu98}.  The techniques
described here and in ref.  \cite{hunt04} open a new window on these
and other quantum optics experiments.

This work was supported by the
Australian Research Council and by the Schwerpunktprogramm 1078 of the Deutsche
Forschungsgemeinschaft and the network of competence QIP of the State
of Bavaria (A8).  U. A. gratefully acknowledges financial support of
the Alexander von Humboldt foundation.


\end{document}